\documentclass{article}
\usepackage{epsfig}
\begin{document}

\title{\bf Pathways to folding, nucleation events and native geometry}

\author{
{\bf Rui D.M. Travasso}\\
Centro de F\'\i sica Te\'orica e Computacional, Faculdade de Ci\^encias, Universidade 
de Lisboa\\ Av. Prof. Gama Pinto 2, 1649-003 
Lisboa, Portugal\\
email: rui@cii.fc.ul.pt\\
\\
{\bf Margarida M. Telo da Gama} \\
Centro de F\'\i sica Te\'orica e Computacional, Faculdade de Ci\^encias, Universidade 
de Lisboa\\ Av. Prof. Gama Pinto 2, 1649-003 
Lisboa, Portugal\\
Departamento de F\'\i sica, Faculdade de Ci\^encias, Universidade 
de Lisboa\\ Campo Grande, Edif\'\i cio C8,
1749-016 Lisboa, Portugal\\
email: margarid@cii.fc.ul.pt\\ 
\\
{\bf Patr\'\i cia F. N. Fa\'\i sca }\\
Instituto de Tecnologia Qu\'\i mica e Biol\'ogica, Universidade Nova de Lisboa\\ 
Av. da Rep\'ublica, EAN 2785-572 Oeiras, Portugal \\
email: patnev@cii.fc.ul.pt, author to whom correspondence should be addressed\\
}

\maketitle

\abstract{\bf{We perform extensive Monte Carlo simulations of a lattice model and the
Go potential to investigate the existence of folding pathways at the level of contact
cluster formation for two native structures with markedly different geometries. 
Our analysis of folding pathways revealed a common underlying folding mechanism, based on nucleation phenomena, for both protein models. However, folding to the more complex geometry (i.e. that with more non-local contacts) is driven by a folding nucleus whose geometric traits more closely resemble those of the native fold. For this geometry folding is clearly a more cooperative process.}}

\section{Introduction} 

Protein folding is the process by which a linear chain of amino acids
{\it spontaneously} acquires a specific three-dimensional native 
structure ~\cite{ANFINSEN}. As pointed out by Levinthal
in the late 1960s a random search of the conformational space for the
global minimum of the free energy (i.e. for the unique native fold) is
not compatible with the biological timeframe of
folding~\cite{LEVINTHAL}. This  raised the hypothesis that folding
might have to occur through an ordered sequence of events (i.e. an
ordered sequence of conformational changes) for the protein to reach 
rapidly its native conformation when starting from the unfolded state. In
other words, kinetic pathways of folding, comprising or not specific
intermediates, were envisaged to explain the timescale of protein
folding~\cite{LEVINTHAL, BALDWIN}. 
 
The discovery in the early 1990s that the 64-residue protein
chymotripsin inhibitor 2 (CI2) folds rapidly with single-exponential
(two-state) kinetics~\cite{JACKSON1} showed  that the existence of
discrete folding intermediates is not a pre-requisite to fold
fast. Indeed, the vast majority of small (with less than 120 amino
acids), single domain proteins  are, like CI2, rapid two-state
folders~\cite{JACKSON2}. Another `simplifying' feature of small
proteins is their topology-dependent folding kinetics; the contact order, CO,
~\cite{PLAXCO}, measuring the average sequence separation of
contacting residues in the native fold, and other related metrics of
native geometry~\cite{GROMIHA, ZHOU} are strongly correlated with
folding rates, suggesting that native topology plays a key role in
determining the folding mechanism.   

A protein engineering method termed $\phi$-value
analysis~\cite{FERSHT1} revealed that CI2 folds via
nucleation-condensation (NC)~\cite{FERSHT2}, a mechanism that was
first observed by Shakhnovich and collaborators in the context of
simulations of lattice proteins~\cite{SHAKHNOVICH}. In the NC
mechanism the formation of a small set of local native bonds, stabilized
by a few non-local native interactions, the so-called folding nucleus,
triggers the rapid emergence of the native fold. Subsequent studies
suggested that NC is possibly the most common folding mechanism
amongst single domain proteins~\cite{NOL}.   

The problem of identifying folding pathways along with the formation of folding nuclei is therefore of the utmost importance in protein chemistry and has been investigated within different frameworks~\cite{TIANA1, HUBNER}. Computer simulations of protein folding and unfolding, both on- and off-lattice, have proved particularly useful in exploring protein folding pathways and mechanisms at different levels of structural detail~\cite{PANDE,MAREK1,TIANA1, TIANA2,CAFLISH,MAREK2,SEELIGER,MAREK3,IRBACK,HUBNER,SUTTO,LEANDRO,
TODD, WEIKL}. For example, at the micro-structural level of contact
formation it was shown that folding is dominated by a well-defined
sequence of events~\cite{TIANA1} and that the sequencing of events
depends primarily on the native geometry as defined by the
CO~\cite{MAREK1}. On the other hand a more recent study, revealed that
the unfolding process of CI2 happens in a highly parallel
fashion~\cite{WEIKL}. At a coarser level of structure defined by contact 
clusters (i.e. secondary structure elements) sequential folding events have
been reported within different simulational frameworks~\cite{MAREK1,WEIKL,IRBACK,CAFLISH,MAREK3}.    

Here we use a lattice model and the G\={o} potential to explore in some detail the folding pathways leading to different native geometries. In particular, we determine the order according to which different sections of the native fold become structured as folding progresses toward the native state. For both geometries there is one section that exhibits a distinctively different folding pattern. Moreover, the timely formation of this particular section determines the most probable folding pathways. By comparison with previous studies, based on specific strategies to identify the folding nucleus, we have confirmed that this unique section, identified through the analysis of the folding pathways, does contain the critical contacts forming the folding nucleus.   
      
This article is organized in the following way. In the next section we describe the model and simulational methods employed, then we present and discuss the results of the simulations, and finally we draw some concluding remarks.

\section{Model and Methods}

\subsection{G\={o} model and simulation details}

We consider a simple three-dimensional lattice model of a protein molecule with
chain length $N$=48. In such a minimalist model amino acids, represented
by beads of uniform size, occupy the lattice vertices and the peptide
bond, that covalently connects amino acids along the polypeptide
chain, is represented by sticks with uniform (unit) length
corresponding to the lattice spacing. \par 
To mimic protein energetics we use the G\={o} model~\cite{GO}. In the
G\={o} model the energy of a conformation, defined by the set of bead
coordinates $\lbrace \vec{r_{i}} \rbrace$, is given by the contact
Hamiltonian   
\begin{equation}
H(\lbrace \vec{r_{i}} \rbrace)=\sum_{i>j}^N
\epsilon \Delta(\vec{r_{i}}-\vec{r_{j}}),
\label{eq:no1}
\end{equation}
where the contact function $\Delta (\vec{r_{i}}-\vec{r_{j}})$, is unity only 
if beads $i$ and $j$ form a non-covalent native contact, i.e., a contact 
between a pair of beads that is present in the native structure, and is zero otherwise. 
The G\={o} potential is based on the idea that the native fold is very 
well optimized energetically. Accordingly, it ascribes equal stabilizing 
energies (e.g., $\epsilon=-1.0 $) to all the native contacts and neutral energies 
($\epsilon =0$) to all non-native contacts. As the G\={o} model has a uniform 
distribution of contact energies the folding dynamics driven by the G\={o}
potential is essentially determined by the structural features of the native fold.\par
In order to mimic the protein's relaxation towards the native state we use a 
Metropolis Monte Carlo (MC) algorithm~\cite{METROPOLIS, MAREK_METHOD1, MAREK_METHOD2} together with the kink-jump move set~\cite{BINDER}. 
To guarantee that the detailed balance condition is satisfied the probability of a certain conformational change must be independent of the conformation adopted by the chain~\cite{MAREK_METHOD1, MAREK_METHOD2}. Therefore, at each MC step, the probability of applying the Metropolis criteria to a particular chain displacement is $0.2/(N+6)$ if the change involves moving one single bead (end-move and corner-flip), or $0.8/(N-3)$ if it involves the simultaneous movement of two beads (crankshafts). 
A MC simulation starts from a randomly generated unfolded conformation and the folding dynamics is monitored by following the evolution of the fraction of native contacts, $Q=q/L$, where $L$ is the number of contacts in the native fold and $q$ is the 
number of native contacts formed at each MC step. The number of MC
steps required to fold to the native state (i.e., to achieve $Q=1.0$)
is the first passage time (FPT) and the folding time is computed as
the mean first passage time (MFPT) of 300 simulations. Folding is studied 
at the so-called optimal folding temperature, $T_{opt}$, the temperature that minimizes the folding time~\cite{OLIVEBERG, JCPSHAKH, CIEPLAK, PFN1}. 

\subsection{Native Geometries}
\begin{figure}
\center{\Large Geometry 1}
$$
\begin{array}{cc}
\epsfig{file=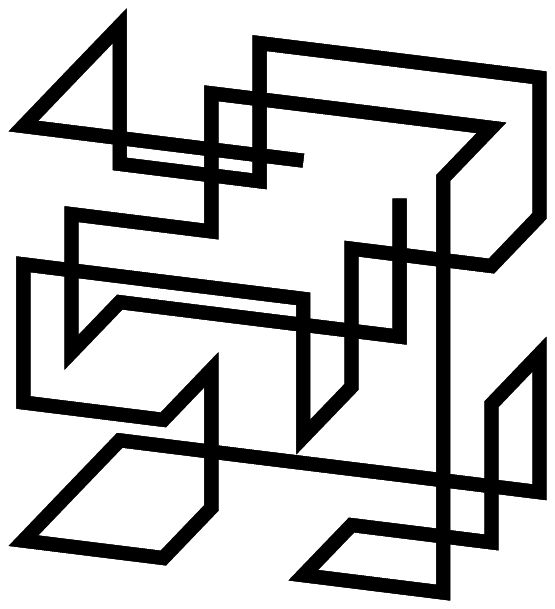, width=7.5cm, angle=180} & \epsfig{file=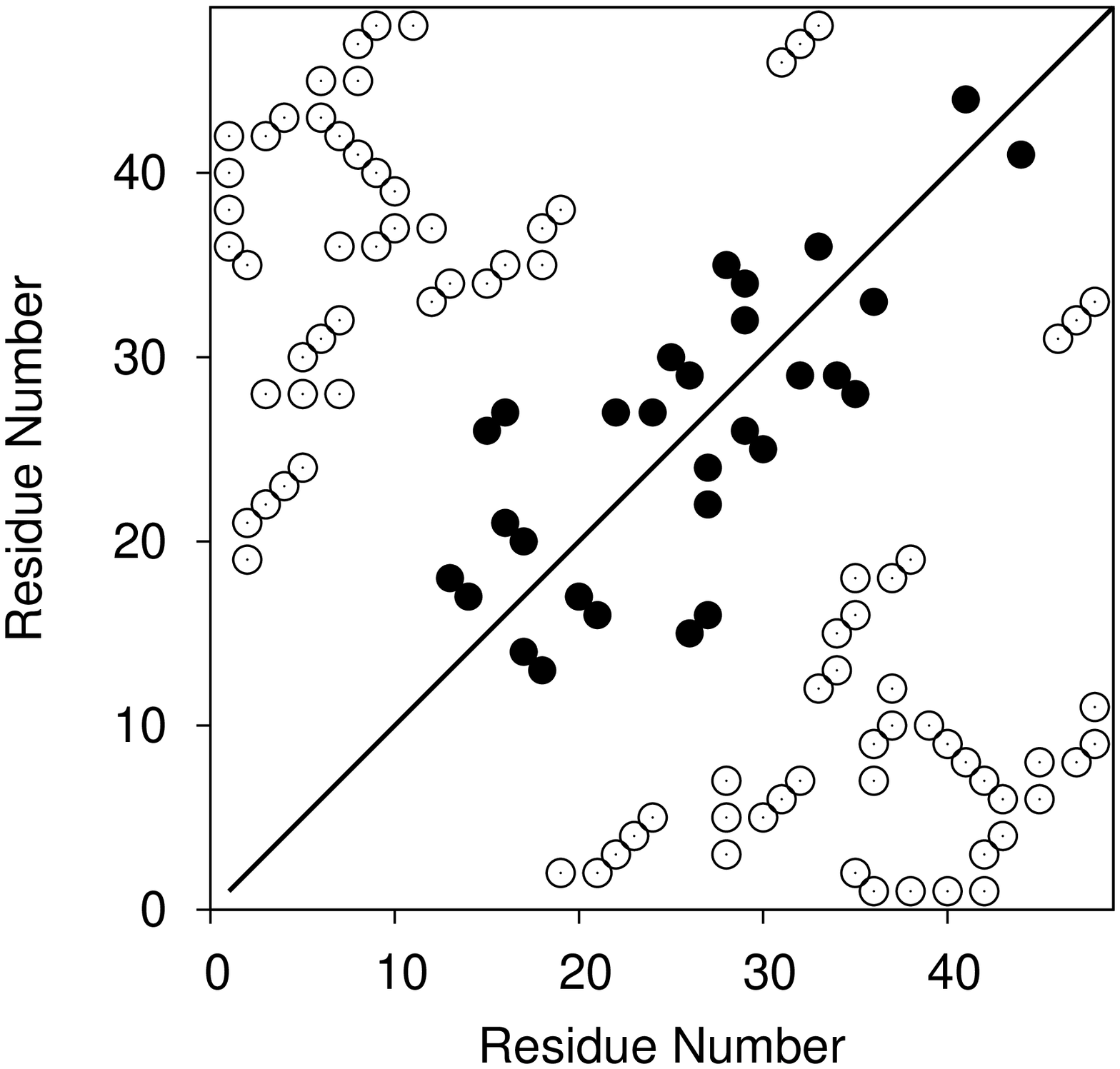,,
  width=5cm, angle=-90}
\end{array}
$$
\center{\Large Geometry 2}
$$
\begin{array}{cc}
\epsfig{file=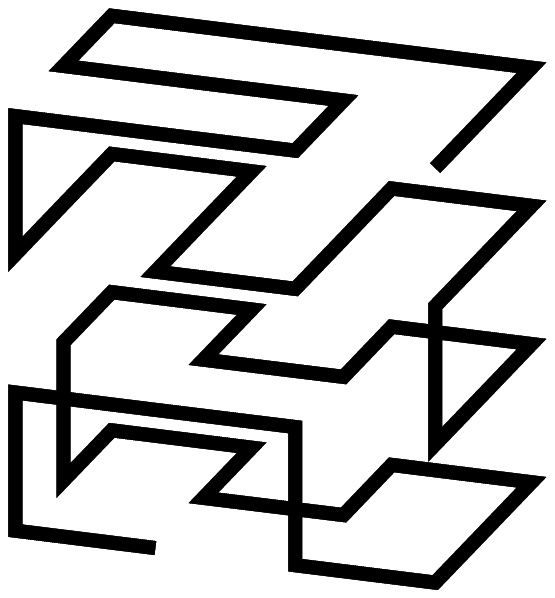, width=7.5cm, angle=180} & \epsfig{file=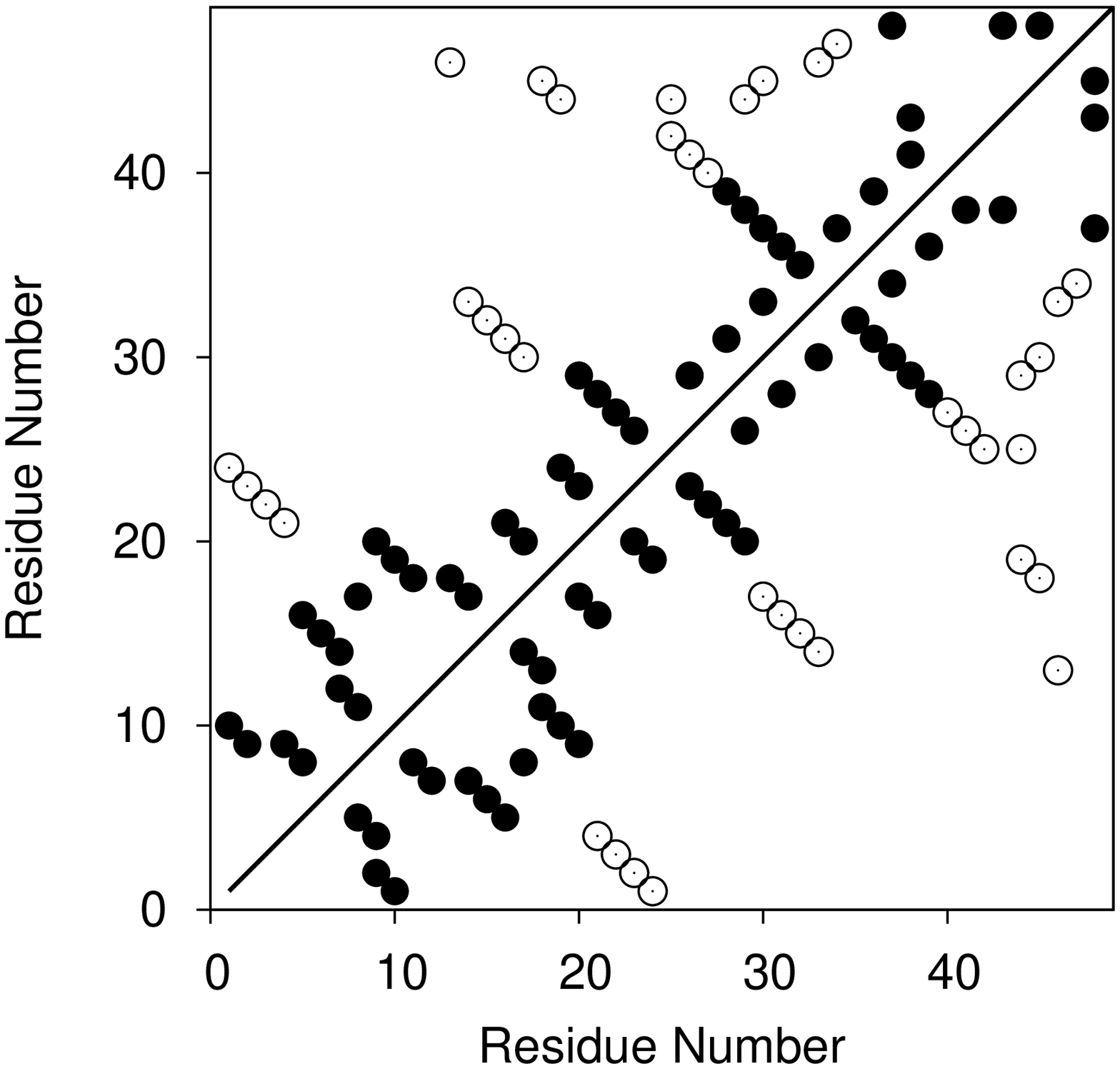,
  width=5cm, angle=-90}
\end{array}
$$
\caption{\label{nativegeo}Three-dimensional representation of Geometry 1 (top, left) and Geometry 2 (bottom, left), and their respective contact maps (right). In the contact maps each circle represents a native contact. Non-local LR contacts are shown in white.}
\end{figure}

In order to explore how native geometry alone drives the folding process,
two native folds (Figure~\ref{nativegeo}), which are amongst the most
complex (Geometry 1) and the simplest (Geometry 2) cuboid
geometries found through lattice simulations of homopolymer
relaxation~\cite{PFN2}, were considered in this study. 
For structures that like ours are maximally compact cuboids with N$=48$
residues, there are 57 native contacts. A non-local contact between
two residues $i$ and $j$ is considered long-range (LR) if its sequence
separation is at least 12 units, i.e. $|i-j| \geq
12$~\cite{GROMIHA}. Geometry 1 is characterized by a large number of
LR contacts while in Geometry 2 the native bonds are predominantly local
(Figure~\ref{nativegeo} and Table \ref{propriedades}). The larger number of LR contacts in
Geometry 1 translates into a large absolute contact order (ACO)~\cite{PLAXCO}. 
\begin{table}
\begin{center}
\begin{tabular}{||c||c|c|c|c||}
\hline
\hline
Geometry & ACO & Fraction LR & $T_{opt}$ & folding time ($\mathrm{x}10^6$ MCS)\\
\hline
1 & 21.4 & 0.74&0.65 & $8.1 \pm 0.5$\\
2 &  10.0 & 0.33&0.66 & $2.3 \pm 0.1$\\
\hline
\hline
\end{tabular}
\end{center}
\caption{\label{propriedades}Absolute contact order (ACO), fraction of
  long-range (LR) contacts, optimal folding
  temperature, $T_{opt}$, and folding time for
  geometries 1 and 2.} 
\end{table}

\subsection{Probability to fold, $P_{fold}$}

\begin{figure}
$$
\begin{array}{cc}
\epsfig{file=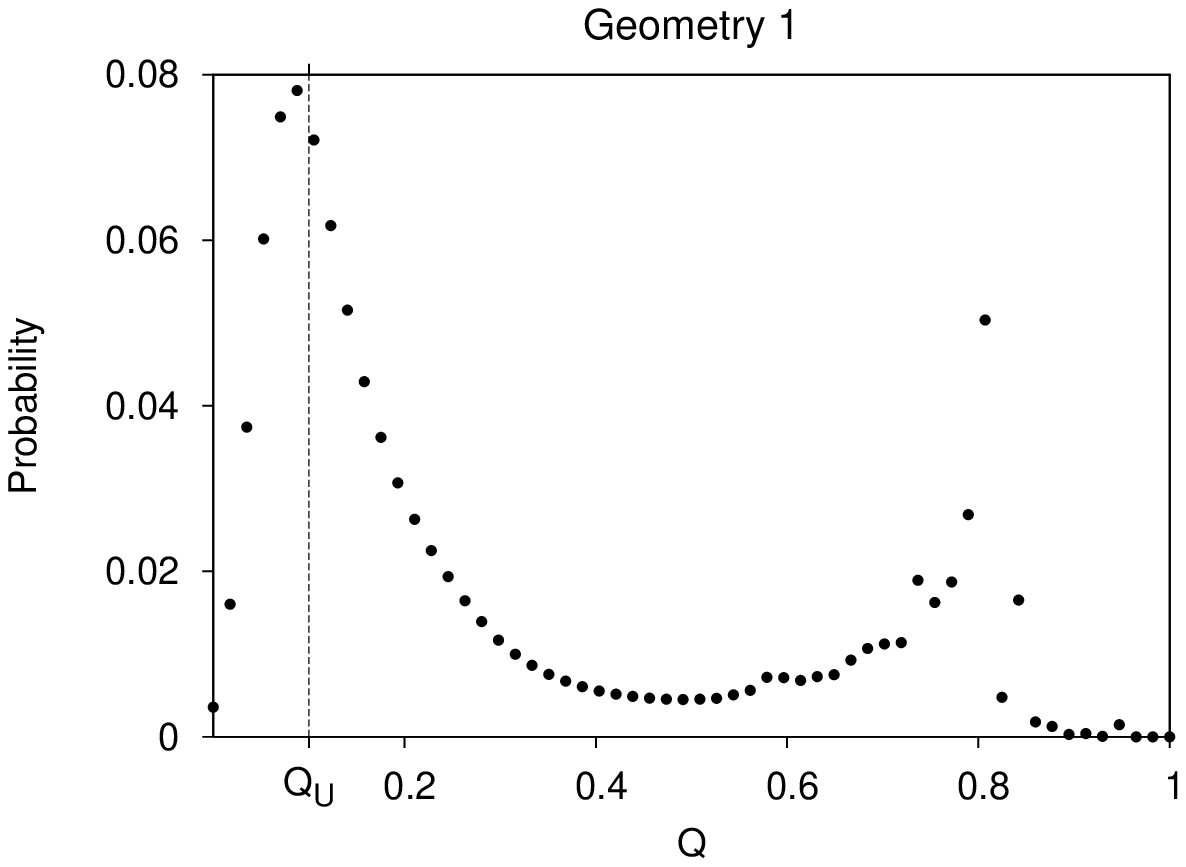, width=7.5cm} & \epsfig{file=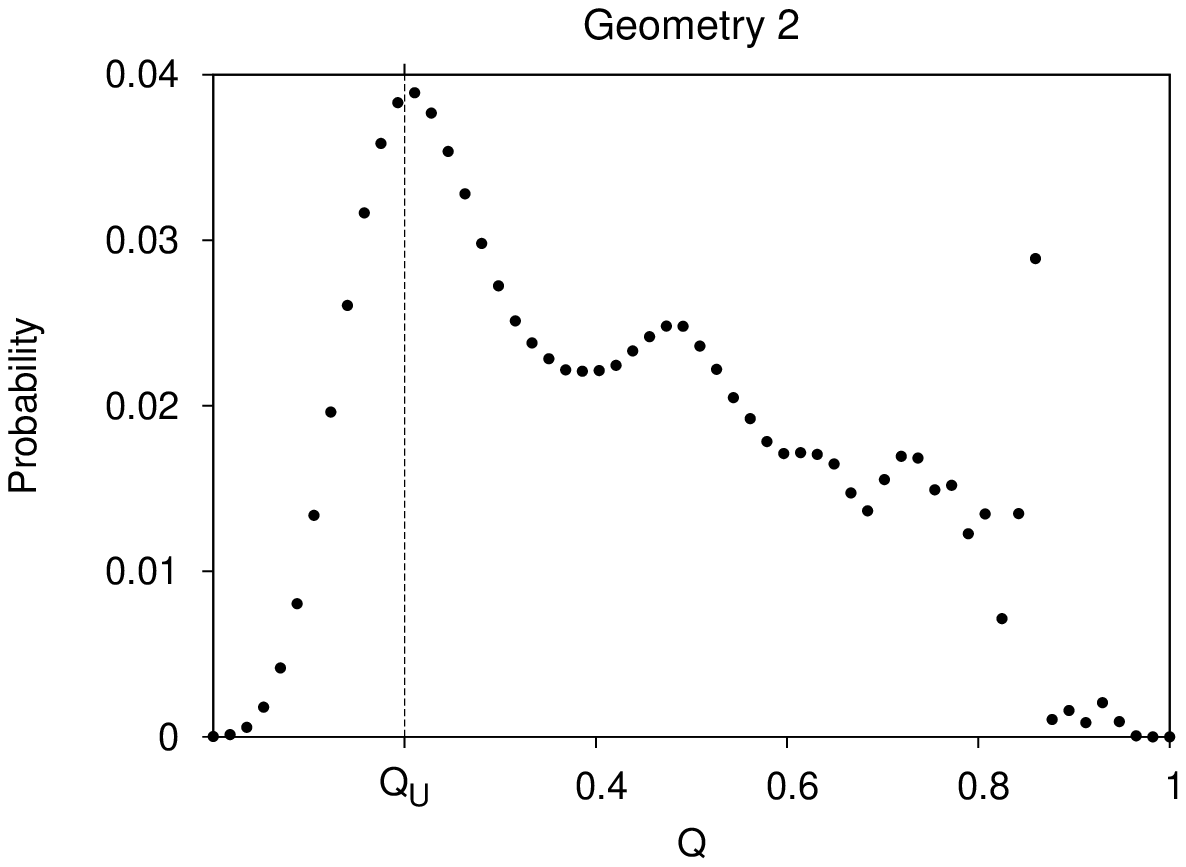,
  width=7.5cm}
\end{array}
$$
\caption{Probability distribution for the fraction of native contacts,
$Q$, for Geometry 1 (left) and Geometry 2 (right) as a function of 
$Q$. A conformation is considered unfolded when $Q < Q_U$.} 
\label{q} 
\end{figure}

The folding probability, $P_{fold}(\Gamma)$, for a conformation
$\Gamma$ is defined as the the fraction of MC runs which, starting from 
$\Gamma$, fold before they unfold~\cite{DU}. To compute
$P_{fold}$ we use an ensemble of 500 MC runs divided into bins of 100 runs.
$P_{fold}$ is firstly computed for each bin and the values thus found are subsequently averaged, and the respective standard deviation evaluated. Each MC run stops when either the native fold or some unfolded conformation is reached. A conformation is deemed unfolded when its total fraction of native contacts, $Q$, is smaller than some cut-off, $Q_{U}$. In order to estimate $Q_{U}$ we compute the probability of finding some fraction of native contacts $Q$ as a function of $Q$ in 200 MC folding runs (Figure \ref{q}). Considerably small $Q$ must necessarily identify unfolded (or denatured) conformations. Indeed, a high-probability peak, centered around the fraction of native contacts $Q_{U}=0.1$,  is readily apparent in the graph reported for Geometry 1 (Figure~\ref{q}, left). Similarly, the highest probability peak appears around $Q_{U}=0.2$ for Geometry 2 (Figure~\ref{q}, right). These fractions of native contacts are relatively low and therefore identify states with minimal residual structure. In this work we use these values of $Q$ to define each geometry's cut-off value $Q_U$.

\section{Exploring the hidden `architecture' within a lattice protein}

In real globular proteins native contacts are {\it clustered} into
the so-called secondary structure elements ($\alpha$-helices,
$\beta$-sheets etc.), which have no direct analogue on the lattice.
Therefore, in this coarse grained representation, there are not
well defined clusters of contacts associated with the secondary
structural elements. Nevertheless, it is possible to identify well defined
clusters of contacts in lattice proteins that form  
well-defined sections of the native fold. We have developed a method
(based on inter-residue contact correlation analysis) that groups
native contacts into distinct protein sections according to whether
their presence is strongly correlated.\par 
 
\subsection{Target conformations}
\label{sect}
\begin{figure}
$$
\begin{array}{cc}
\epsfig{file=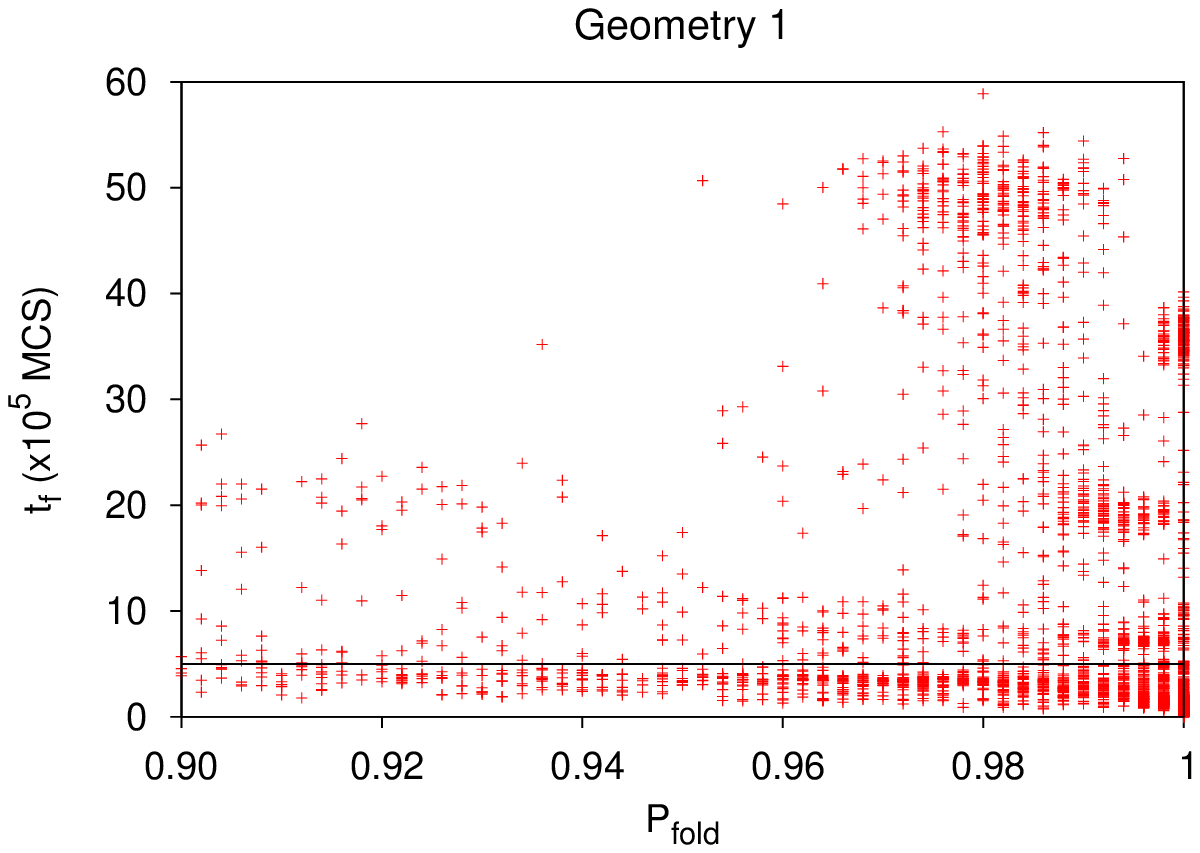, width=7.5cm} & \epsfig{file=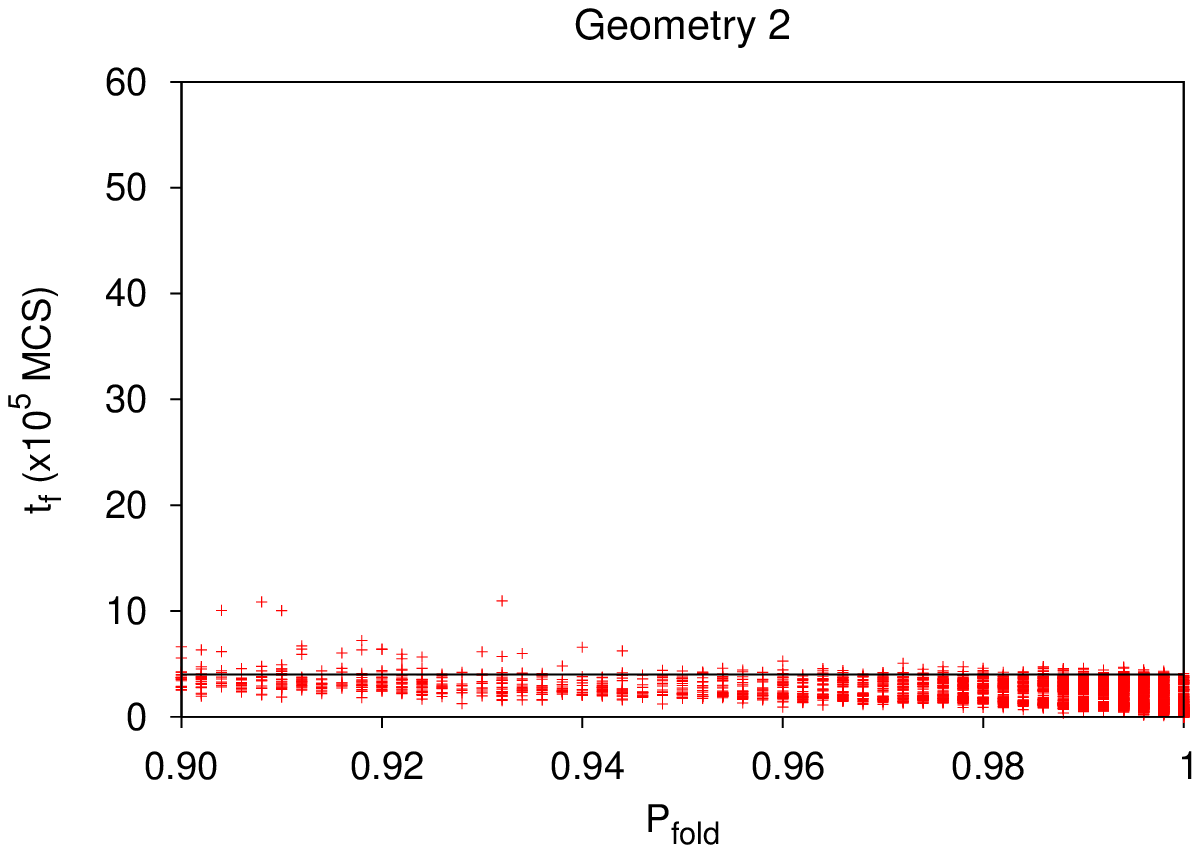,
  width=7.5cm}
\end{array}
$$
\caption{Time-to-fold, $t_{f}$, as a function of the reaction coordinate, $P_{fold}$. 
Long lived trapped states are observed in Geometry 1 (left) at very high
$P_{fold}$, but are absent in Geometry 2 (right). To measure $t_f$ of each conformation 
we considered only folding events in which the protein folded before unfolding. $t_f$ is the mean time-to-fold averaged over these folding events. 
The horizontal black lines indicate the cut-off times below which a
conformation is committed to fold. For Geometry 1 and 2 there are respectively 4724 and 4162 conformers with $P_{fold} \geq 0.9$.
\label{txp}}   
\end{figure}

The first step in the proposed procedure is that of selecting an
ensemble of appropriate target conformations. These must be
considerably native-like and, most importantly, committed to fold.
In order to find such productive conformers we ran 8000 MC simulations
for each model geometry and sampled a conformation from each independent MC 
run when folding was near completion (i.e. at a time close to the run's FPT). 
Conformations thus selected are dynamically uncorrelated
and provide a sample of statistically independent elements. For every
conformation we have computed $P_{fold}$, along with its
standard deviation. The time-to-fold, $t_f$, was then measured  
for conformations with $P_{fold} \geq 0.9$.
To compute $t_f$ we have only considered MC runs where the proteins  
to fold before they unfold. {\it A priori} one would expect such high-$P_{fold}$ conformations to be kinetically very close to the native state. However, for Geometry 1, a plot of the dependence of $t_f$ on the folding probability reveals the existence of many
conformations with $P_{fold}>0.98$ that find the native state in a timeframe comparable with that observed in simulations starting from random coil-type conformations 
(Figure \ref{txp}, Table \ref{propriedades}).
These are trapped states (i.e. off-pathway to folding) and are eliminated from the initial sample. Indeed, the vast majority (i.e. 73\%) of high $P_{fold}$ conformers find the native state in less than 6\% of the
folding time. These conformations are on the folding pathway and can be used to 
cluster native bonds and shed light into the existence of putative protein sections. The mean fraction of native contacts in these conformations is $<Q>=0.73$ and, on average, they differ by 10.31 native contacts.\par  

For Geometry 2, 95\% of conformations with high
$P_{fold}>0.9$ rapidly find the native state in less than  17\% 
of the folding time. On average they differ by 9.18 contacts and, like in 
Geometry 1, their mean fraction of native contacts is $<Q>=0.73$. 

\subsection{Inter-residue contact correlation analysis reveals distinct protein sections}
\label{secant}
\begin{figure}
\center{\Large Geometry 1}
\begin{center}
\epsfig{file=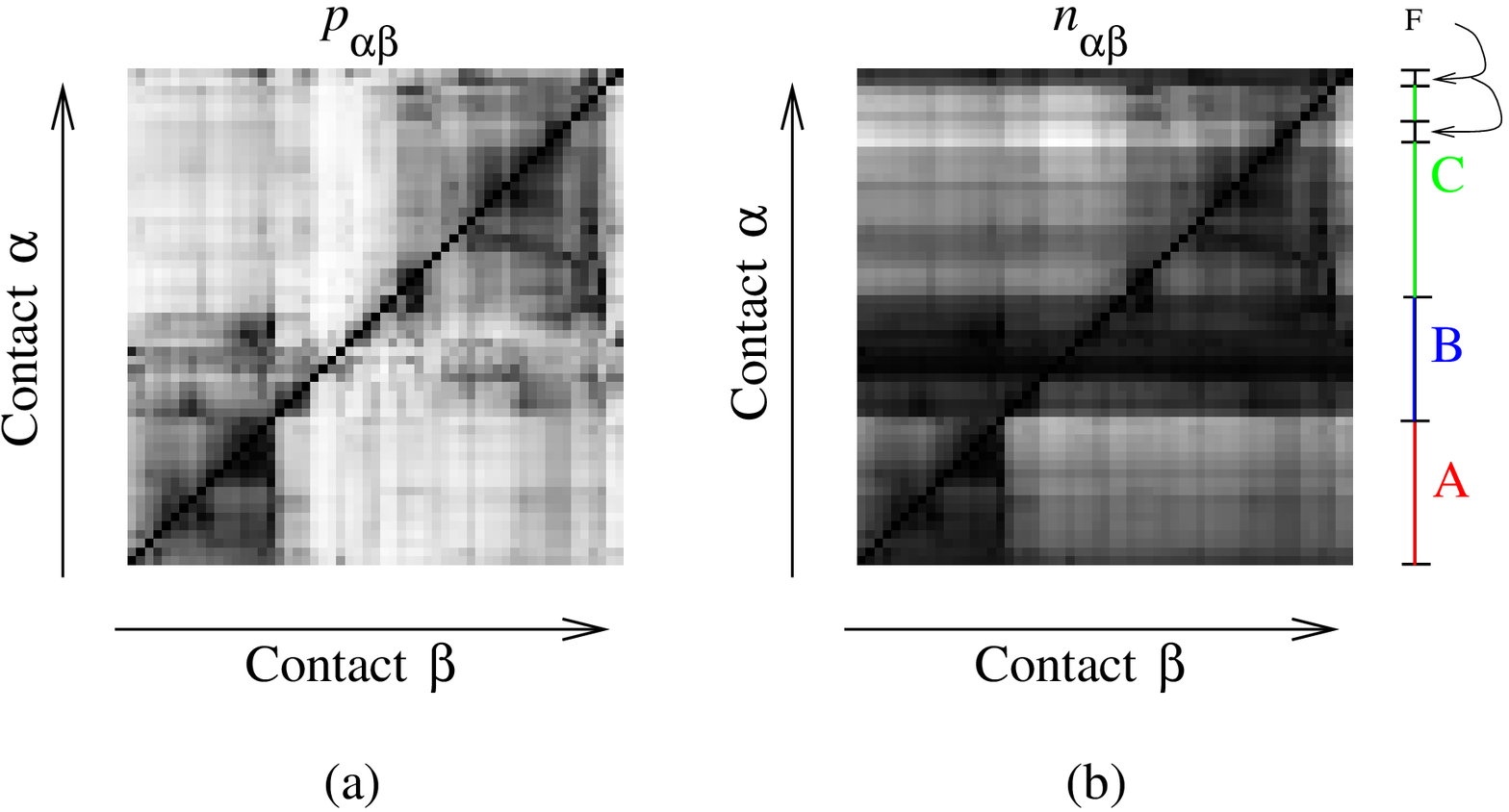, width=12cm}
\end{center}
\center{\Large Geometry 2}
\begin{center}
\epsfig{file=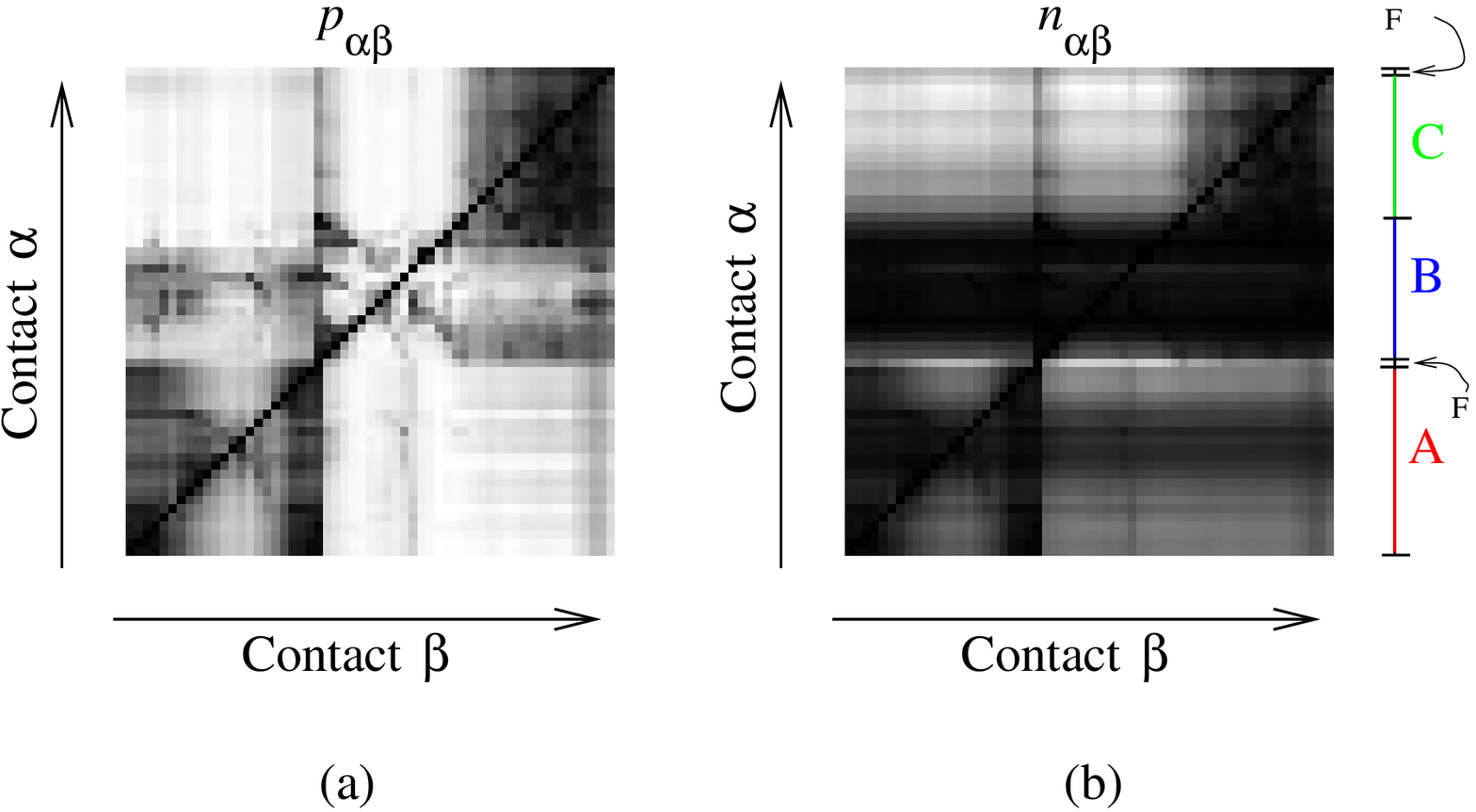, width=12cm}
\caption{Density plots of the probability (left column) and fraction of
  conformations (right column) where contact $\beta$ is present and $\alpha$ is not for 
  Geometry 1 (top) and Geometry 2 (bottom). Native contacts are ordered 
  according to their relative values of $C_{\alpha\beta}$ (the order is the same for the $p_{\alpha\beta}$ and $n_{\alpha\beta}$ plots). The groups of contacts forming   
  sections A, B, and C are identified. 
  Contacts that were not assigned to any section (`free' contacts) are
  identified by the letter F. The range of $p_{\alpha\beta}$ lies between 0 (black) and 1 
  (white), while $n_{\alpha\beta}$ varies between 0 (black) and 0.54 (white) in Geometry 1 and between
  0 (black) and 0.64 (white) in Geometry 2. 
\label{sq}}
\end{center} 
\end{figure}

In conformations with high $P_{fold}$ that are committed to fold 
one expects that protein sections, comprising groups of correlated native bonds, 
will be formed with considerably high probability. We say that two
native contacts $\alpha$ and $\beta$ are correlated i.e., that they belong to
the same section, if
(i) they have similar probabilities of being present when an arbitrary third
contact $\gamma$ is not, and (ii) the probability of contact $\gamma$ being
present if contact $\alpha$ is not is similar to the probability of $\gamma$
being present if contact $\beta$ is absent. Formally, 
conditions (i) and (ii) may be quantified 
by correlation between $\alpha$ and $\beta$, $C_{\alpha\beta}$, defined as 
\begin{equation}
C_{\alpha\beta}=\frac{\sum_{\gamma\neq \alpha,\beta} n_\gamma (p_{\gamma\alpha}-p_{\gamma\beta})^2 +
  \frac{n_\alpha+n_\beta}{2} \sum_{\gamma\neq \alpha,\beta}
  (p_{\alpha\gamma}-p_{\beta\gamma})^2}{(L-2)\frac{n_\alpha+n_\beta}{2} + \sum_{\gamma\neq \alpha,\beta} n_\gamma}
\label{C}
\end{equation}
being $\ll 1$. In the expression above $p_{\alpha\gamma}$ is the conditional
probability of finding contact $\gamma$ if contact $\alpha$ is not present,
$n_\alpha$ is the number of conformations in the sample 
where contact $\alpha$ is not present, and $L=57$ is the total number of native
contacts. The error associated with $p_{\alpha\gamma}$ is of the
order of $1/\sqrt{n_\alpha}$. 
Therefore the weight of each averaged term in
(\ref{C}) of either $n_\gamma$ or $(n_\alpha+n_\beta)/2$ implies that $C_{\alpha\beta}$ is 
determined by the terms which are measured with the highest accuracy   
(this is an important point since the measurement error associated 
with the difference between probabilities $p_{\gamma\alpha}$ and $p_{\gamma\beta}$ increases as $n_\gamma$ decreases).   
Using equation (\ref{C}) the correlation between pairs of native
contacts $\alpha$ and $\beta$ is computed in the ensembles of target conformations 
selected for  Geometry 1 and Geometry 2, and native contacts are ordered according
to their relative correlations in the following way: starting with an
arbitrary contact, say contact $0$, contact $1$ is the one with the
lowest $C_{01}$ {\em i.e.}, the contact that is the most strongly
correlated with contact $0$, contact $2$ is that with the lowest
$C_{12}$, and so on. This ordering method sheds light on existing
protein sections since it block diagonalizes the contact matrix, $\bf {C}$.
Indeed, density plots for the probability that  contact $\alpha$ is present if
contact $\beta$ is not, $p_{\alpha\beta}$, and for the
fraction of conformations in the sample satisfying
the same condition, $n_{\alpha\beta}$, reveals the existence of 
three protein sections namely, section A, B and C, in the two 
model proteins (Figure \ref{sq}).  

In both geometries, contacts within sections A and C are strongly
correlated. This is shown by the low probability (i.e. dark) 
squared spots located along the diagonal in the 
$p_{\alpha\beta}$ and in the $n_{\alpha\beta}$ ordered matrices. In the $p_{\alpha\beta}$ plot,  such
well-defined regions indicate that when a contact belonging to 
A (or C) is {\it not} formed, any other contact in A (or C) 
has a considerably low
probability of being formed (Figure \ref{sq}(a)). Correspondingly, the
darker squares identifying sections A and C in the $n_{\alpha\beta}$ matrices show
that for any pair of contacts within those sections, there is a small
number of conformations in which one of the contacts in the pair 
is formed while the other is not (Figure \ref{sq}(b)). 

In the $p_{\alpha\beta}$ and $n_{\alpha\beta}$ density plots the brighter spots
located in the  matrices' off-diagonal indicate that contacts
belonging to C (or A) can be formed with a relatively high probability,
when a contact in A (or C) is missing. Hence, we conclude that 
the target conformations have either A or C formed.

Contacts in section B behave differently from those in sections 
A and C as they are {\it always} present with high-probability.
This is shown by the existence of the white vertical bar in the $p_{\alpha\beta}$ density 
plot (Figure \ref{sq}(a)) and the dark (and homogeneous) horizontal band that 
spans the vertical axis in the $n_{\alpha\beta}$ matrices (Figure \ref{sq}(b)).
The white spots on the diagonal in the $p_{\alpha\beta}$ matrices
indicate that, by contrast to contacts in sections A and C, when one contact
within B is missing, other contacts within B may still be formed with high
probability.

Some contacts are located at the boundaries of the identified
sections. The correlation between their presence and other contacts' presence does not fit the correlation patterns found for sections A, B or C. For this reason we decide not to assign them to any section and denote them by {\it free} contacts. There are five free bonds in Geometry 1 (namely,  4-23, 5-24, 12-33, 13-34 and 25-30) and two free bonds (2-9 and 13-46) in Geometry 2.

\subsection{Section's geometric traits}

\begin{figure}
\center{\Large Geometry 1}
$$
\begin{array}{cc}
\epsfig{file=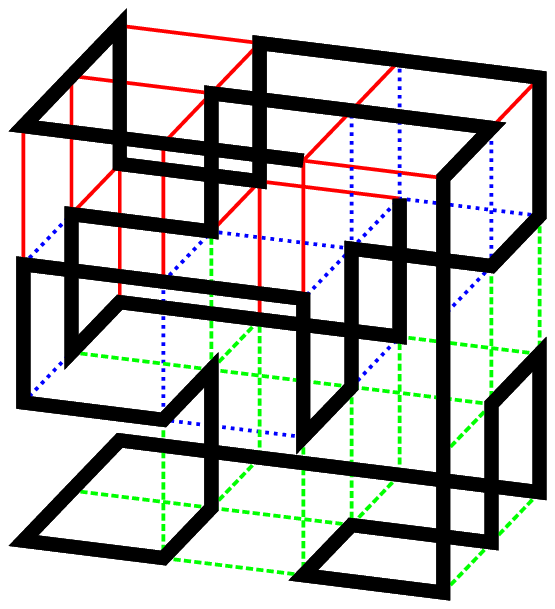, width=7.5cm, angle=180} & \epsfig{file=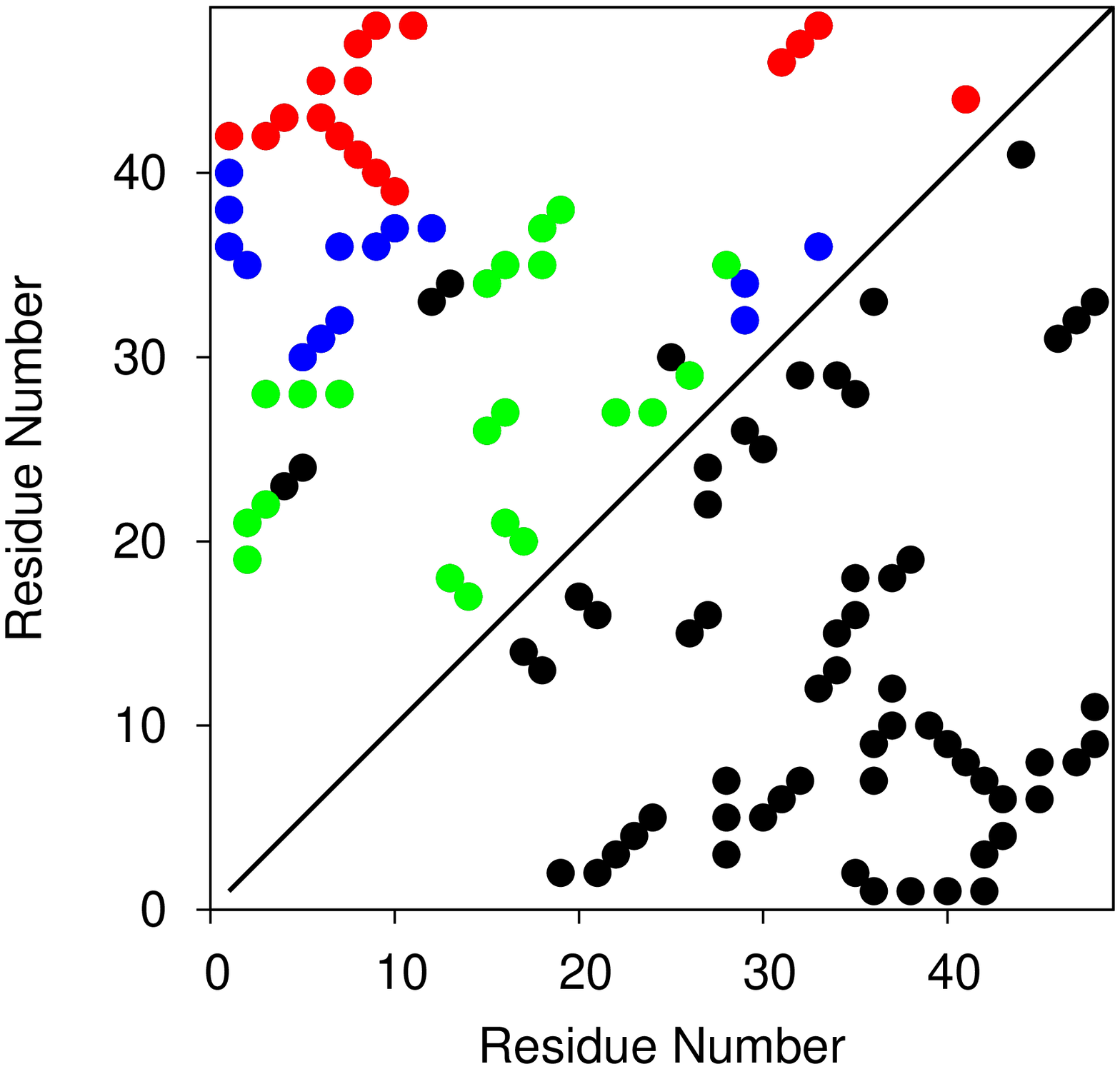,
  width=5cm, angle=-90}
\end{array}
$$
\center{\Large Geometry 2}
$$
\begin{array}{cc}
\epsfig{file=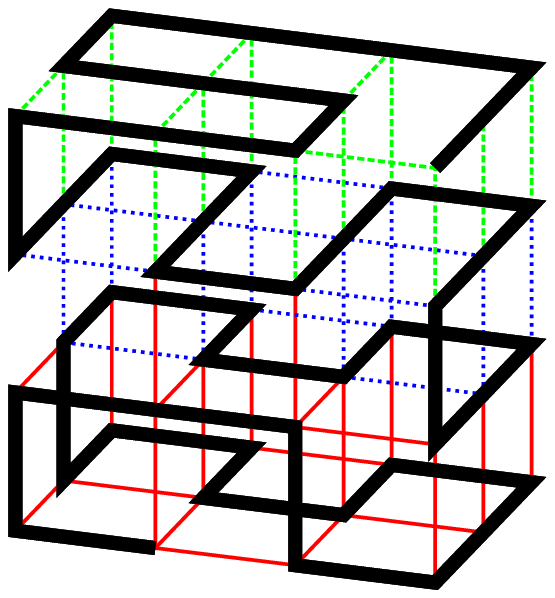, width=7.5cm,angle=180} & \epsfig{file=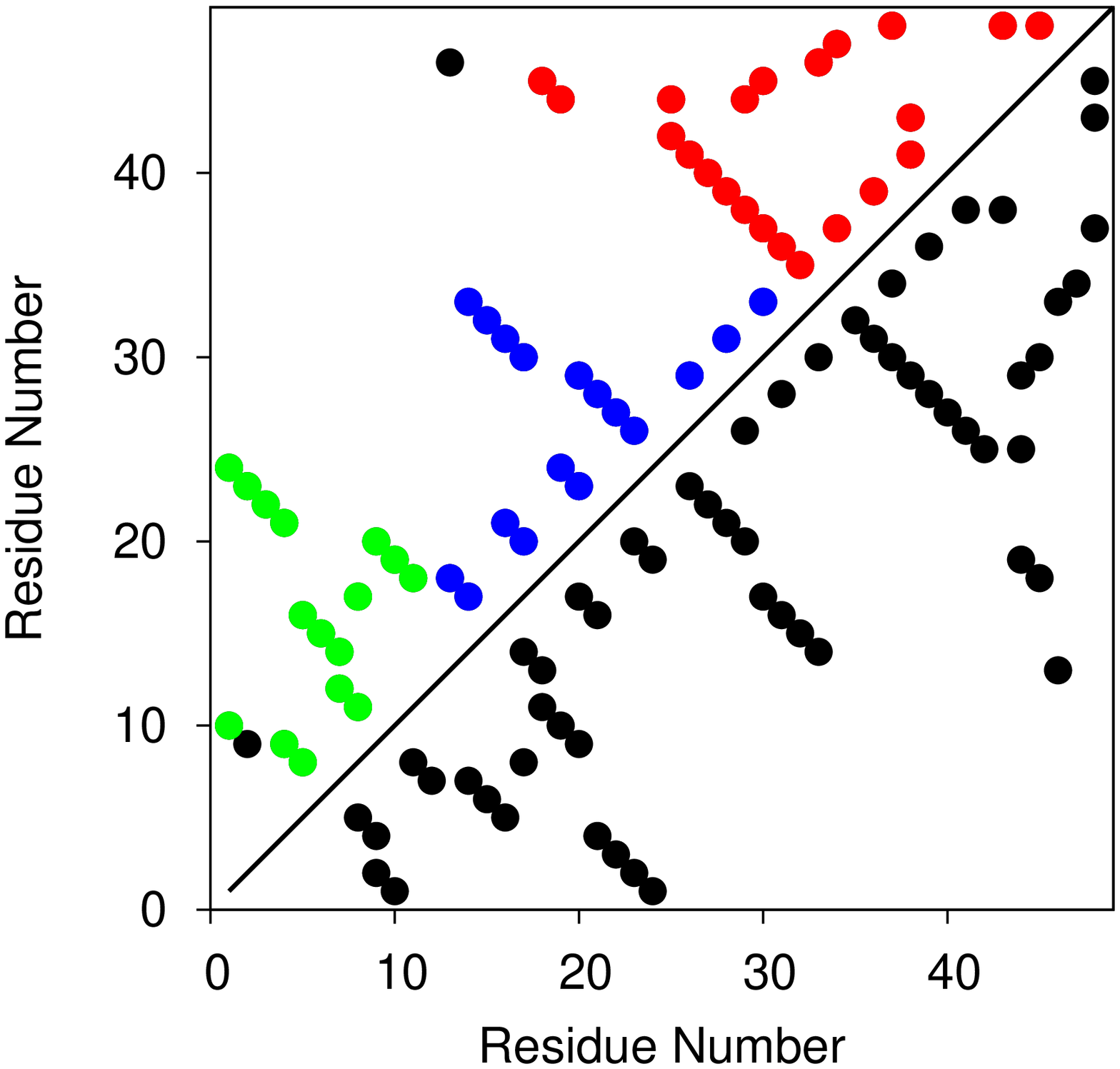,
  width=5cm,angle=-90}\\
\mathrm{(a)} & \mathrm{(b)}
\end{array}
$$
\caption{Protein sections identified for Geometry 1 (top row) and
  Geometry 2 (bottom row). Native contacts forming sections A, B and C
  are respectively colored red, blue and green, in the three
  dimensional representations (left) and contact maps (right). Note that the proteins sections identified as groups of correlated native bonds are grouped together in the protein's three
dimensional native structure.
  \label{g}}  
\end{figure}

The protein sections thus identified as clusters of strongly correlated native
bonds form well defined, separate parts in the native fold (Figure \ref{g}). 
Indeed, clusters of strongly correlated bonds are grouped together in the 
protein's three dimensional representation. 
The structural characterization of each individual section is reported in Table \ref{proper}. 
\begin{table}
\begin{center}
\begin{tabular}{||c|c||c|c|c||}
\hline
\hline
\multicolumn{2}{||c||}{Name} & Number of Contacts&  Fraction LR &ACO \\
\hline
\hline
& Section A & 17 & 0.94 &30.8\\
\cline{2-5}
Geometry 1 & Section B & 14 &0.79 &24.1 \\
\cline{2-5}
& Section C & 21 &0.52 &13.0\\
\hline
\hline
& Section A & 22 & 0.45& 10.9\\
\cline{2-5}
Geometry 2 & Section B & 17  & 0.24& 7.1\\
\cline{2-5}
& Section C & 16 & 0.25 & 10.5\\
\hline
\hline
\end{tabular}
\end{center}
\caption{\label{proper} Number of native bonds forming each protein section, absolute contact order (ACO) and fraction of long-range (LR)contacts of each protein section. 
}
\end{table}

In Geometry 1 the three sections are geometrically different. In
section A, all contacts but one are long-range and link residues located in opposite ends of the chain. On the other hand, about 50\% of the native bonds in section C, are local. They connect residues in the middle of the chain 
(between residues 17 and 34). Contacts forming section B link residues located in the middle of the chain to residues located in either end of the chain. Interestingly, the geometric features of section B reflect those of the overall native fold.
In Geometry 2, on the other hand, the three sections are highly geometrically similar, being formed essentially by local bonds. 

\section{Folding pathways}

A folding pathway is an ordered sequence of events (i.e. of conformational changes) 
observed along the time coordinate. In this section we investigate if
the previously identified protein sections become structured by following
some preferential order, and how such ordering preferences depends on
native geometry. In other words, we investigate the existence of
folding pathways at the macro-structural level of section formation,
and how the latter depend on the native fold geometry.    
In order to do so, the fraction of native contacts in each section,
$Q_S$, is monitored during each folding event.  A
section is considered folded from the time when its fraction of native
contacts $Q_S$ reaches $1.0$ until it decreases below a certain threshold
$Q_S^U$. In other words, the time at which a section folds is the smallest time $t_S$  such that
$Q_S=1.0$ at  time $t_S$ and $Q_S \geq Q_S^U$ at times larger than $t_S$.
 
A priori,  the threshold $Q_S^{U}$ could be section specific. 
However, different values of $Q_S^{U}$ were tested for both proteins
and the results reported hereafter are robust to changes in the exact value 
of this threshold. Therefore, and for the sake of simplicity, each section's $Q_S^U$ was set equal to the cut-off $Q_U$ used previously to determine the whole protein's unfolded state.

We consider 5000 folding events. For each one the times $t_S$ at which
each section folds are recorded and the corresponding folding pathway
is identified. The probability of observing specific pathways is then computed 
(Table \ref{sfo}).

\begin{table}
\begin{center}
\begin{tabular}{||c|c|c||c||c||@{\hspace{.5cm}}||c|c|c||c||c||}
\hline
\hline
\multicolumn{5}{||c||@{\hspace{.5cm}}||}{Geometry 1}
&\multicolumn{5}{|c||}{Geometry 2} \\
\hline
\hline
First & Second & Third & Prob. &Time& First & Second & Third & Prob.&Time\\
A & B & C & 0.28 & 2.2 $\pm$ 0.3&A & B & C & 0.40 & 4.0 $\pm$ 0.2\\
C & B & A & 0.26 & 1.8 $\pm$ 0.2&C & B & A & 0.31 & 5.2 $\pm$ 0.2\\
B & A & C & 0.16 & 1.7 $\pm$ 0.1&B & A & C & 0.13 & 3.4 $\pm$ 0.1\\
B & C & A & 0.11 & 0.9 $\pm$ 0.2&B & C & A & 0.12 & 2.7 $\pm$ 0.1\\
A & C & B & 0.04 & &A & C & B & 0.00&\\
C & A & B & 0.00 & &C & A & B & 0.00&\\
\hline
\hline
\end{tabular}
\end{center}
\caption{Folding pathways at the macro-structural level of section
  formation (showing the first, second and third section to fold) and their relative probabilities of occurrence. The probabilities do not add to one,
  since there are some events in which two sections fold simultaneously. The average 
  time  elapsing between the formation of the first section and the formation of the last section in each  pathway is given in units of 100000 MCS.\label{sfo}}  
\end{table}

Interestingly  the most probable folding pathways are those in 
which section B is the second to fold.  Structurally, this preference translates into  
folding starting  either at the top or at the bottom of the native structure followed by the
consolidation of the structure's middle `layer'  (Figure \ref{g}). The next most 
probable pathways are those where B folds first and, for both geometries, the probability that
B folds last is vanishingly small. These observations suggest that in either case it is
the folding of section B that determines the probability of a folding pathway.  
We disregarded the folding events in which two sections fold simultaneously 
(i.e. in the same MC step) as they result from the discretization of time and space imposed 
by the lattice.

For the most probable folding pathways we measured the time elapsing between the 
formation of the first section and the emergence of the native structure. For both geometries
the shorter time intervals are observed when section B folds first. However, 
these time intervals are  systematically larger in the folding of Geometry 2. Here, and once 
the first section is completely formed, the protein takes on average 25\% of the folding time to achieve the native state if it follows the slowest pathway. For Geometry 1 the equivalent time interval is just 2.5\% of the overall folding time. This feature is particularly interesting because Geometry 2 folds faster than Geometry 1 (Table\ref{propriedades}).

\section{Section formation as a function of the folding probability}

\begin{figure}
\center{\Large Geometry 1}
$$
\begin{array}{ccc}
\epsfig{file=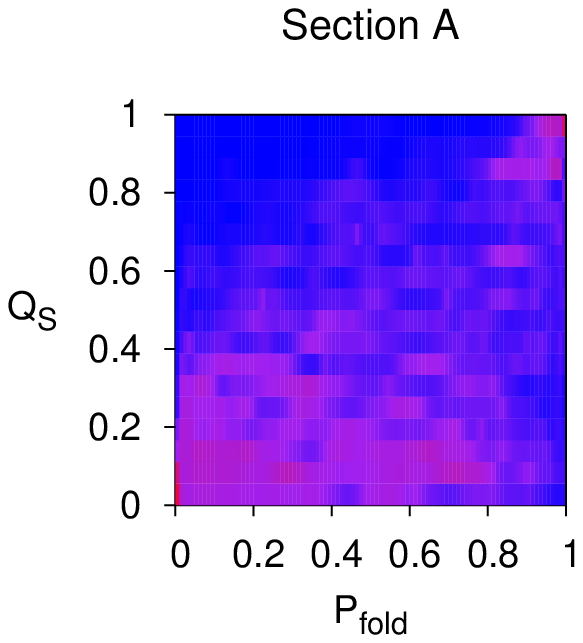, width=5cm} & \epsfig{file=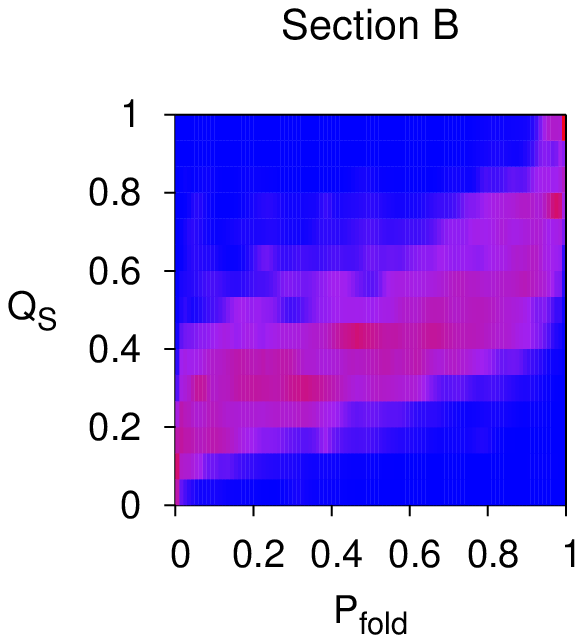,
  width=5cm} & \epsfig{file=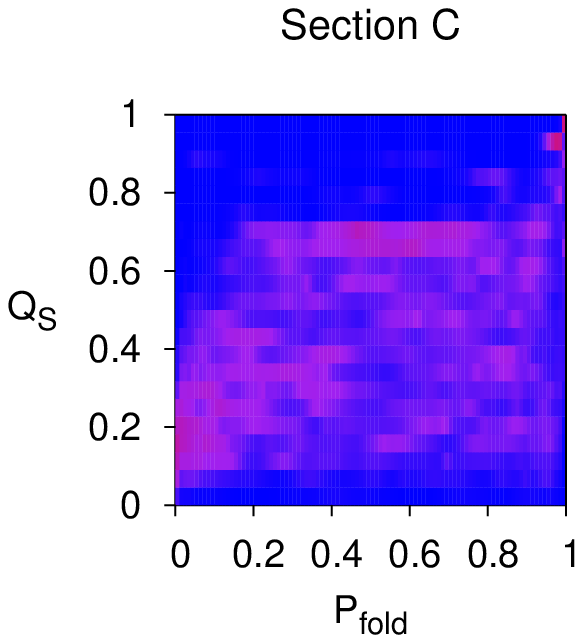, width=5cm}
\end{array}
$$
\center{\Large Geometry 2}
$$
\begin{array}{ccc}
\epsfig{file=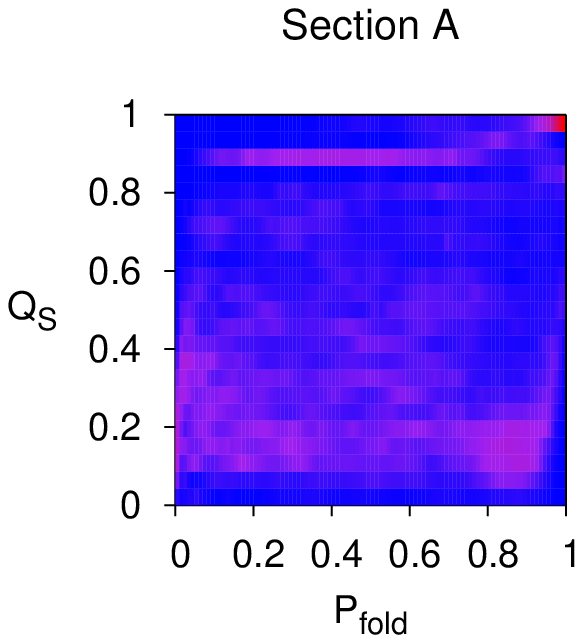, width=5cm} & \epsfig{file=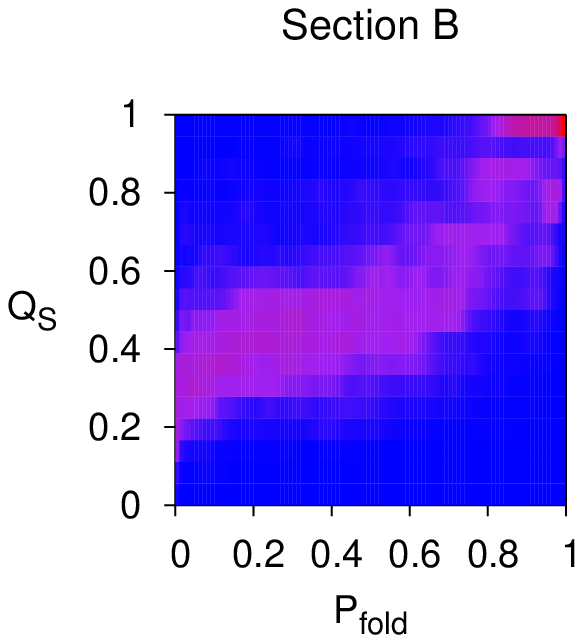,
  width=5cm} & \epsfig{file=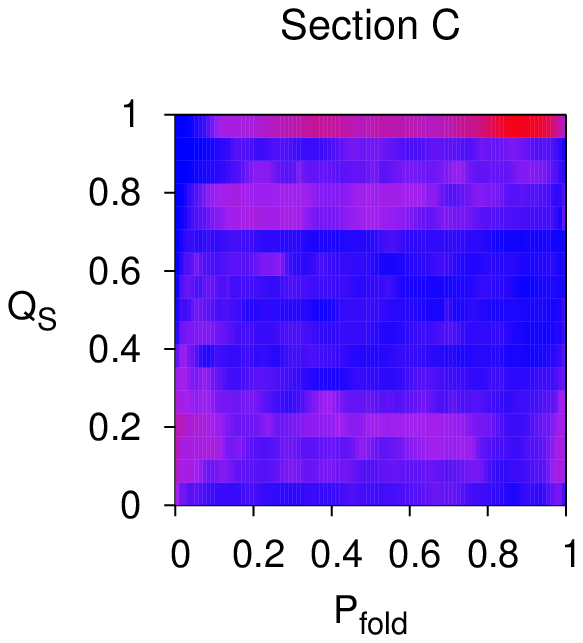, width=5cm}\\
\mathrm{(a)} & \mathrm{(b)}& \mathrm{(c)}
\end{array}
$$
\caption{Density plots of the probability for having a certain $Q_S$
  as a function of $P_{fold}$ for the sections A, B and C in Geometry
  1 (top) and Geometry 2 (bottom). 
 \label{sqq}} 
\end{figure}

Here we analyze the folding progression of individual sections as a function of the probability to fold, $P_{fold}$. In other words, we investigate how the different 
sections of the protein become structured, i.e., how their fraction of native bonds, $Q_S$,
evolves along the folding reaction. In order to do so, two ensembles, each comprising 8000 conformations, were considered for each native geometry and the folding probability of each conformation evaluated 
(section \ref{sect})
\footnote{The standard deviation 
 $\sigma_{P_{fold}}$ was also measured.
Hence, the probability for a conformation $\Gamma$
to have some $P_{fold}$ is considered to be given by the Gaussian distribution with
average $P_{fold}(\Gamma)$  
and standard  deviation $\sigma_{P_{fold}}(\Gamma)$. These Gaussian
distributions are 
used as weighting terms for calculating the probabilities of having 
a section with fraction of native bonds $Q_S$ as a function of
$P_{fold}$}. 
The probabilities of having a section with fraction of native bonds $Q_S$ as 
a function of $P_{fold}$ are shown as density plots in Figure \ref{sqq}.

We start with the analysis of Geometry 1. Here, section
A is essentially unfolded for the most part of the folding
reaction. Indeed, up to $P_{fold} \sim 0.8$, the most probable
conformations are those with fraction of native bonds $Q_A\sim0.1$, and 
it is only when folding is near completion that the probability to find
A folded or close to folded (i.e., with $Q_A>0.9$) is non-zero. 
Due to its local nature, bonds in section C can break and form more easily
than in other sections where non-local bonds abound. It is perhaps for this
reason that $Q_C$ distributes rather uniformly in the range $0.1<Q_C<0.75$
up to late folding stages (i.e., up to $P_{fold}\sim  0.8$,). It is only when 
$P_{fold}>0.9$ that there is a significant group of conformations with more 
than 90\% of section C folded.  

While there is not a correspondence between the behavior of 
$P_{fold}$ and that of the fraction of native contacts formed 
in A and C -- in the sense that higher (lower) $P_{fold}$ does not 
necessarily imply higher (lower) $Q_S$ -- for section B, on the other hand, 
at high $P_{fold}$,  $Q_B$ is on average high, while early on in
folding (at low $P_{fold}$) section B is essentially unfolded.
Therefore, an increase in $P_{fold}$ typically leads to an increase in
$Q_B$, suggesting that the folding of section B acts as a driver for 
the folding of the whole protein.  

For Geometry 2 the folding scenarios of sections A and C are rather distinct 
from those found in the more complex Geometry 1. Indeed, for Geometry 2, the probability 
of finding sections A and C with fraction of native bonds $Q_S$ is strongly bimodal for any $P_{fold}$. This means that at any stage of the folding reaction it is possible to find 
conformations with either A or C almost folded (peak at high $Q_S$)
and others where A and C are very little structured (peak at
low $Q_S$). This observation agrees with our previous findings
regarding the most probable folding pathways, where folding initiates at 
A (and C folds last) or, conversely, it starts at C (and A folds last).  
However, as with Geometry 1, the fraction of native bonds of section B 
increases with $P_{fold}$ and when it achieves some critical value, 
it becomes large enough to prompt folding of sections A or C.

\begin{figure}
$$
\begin{array}{cc}
\epsfig{file=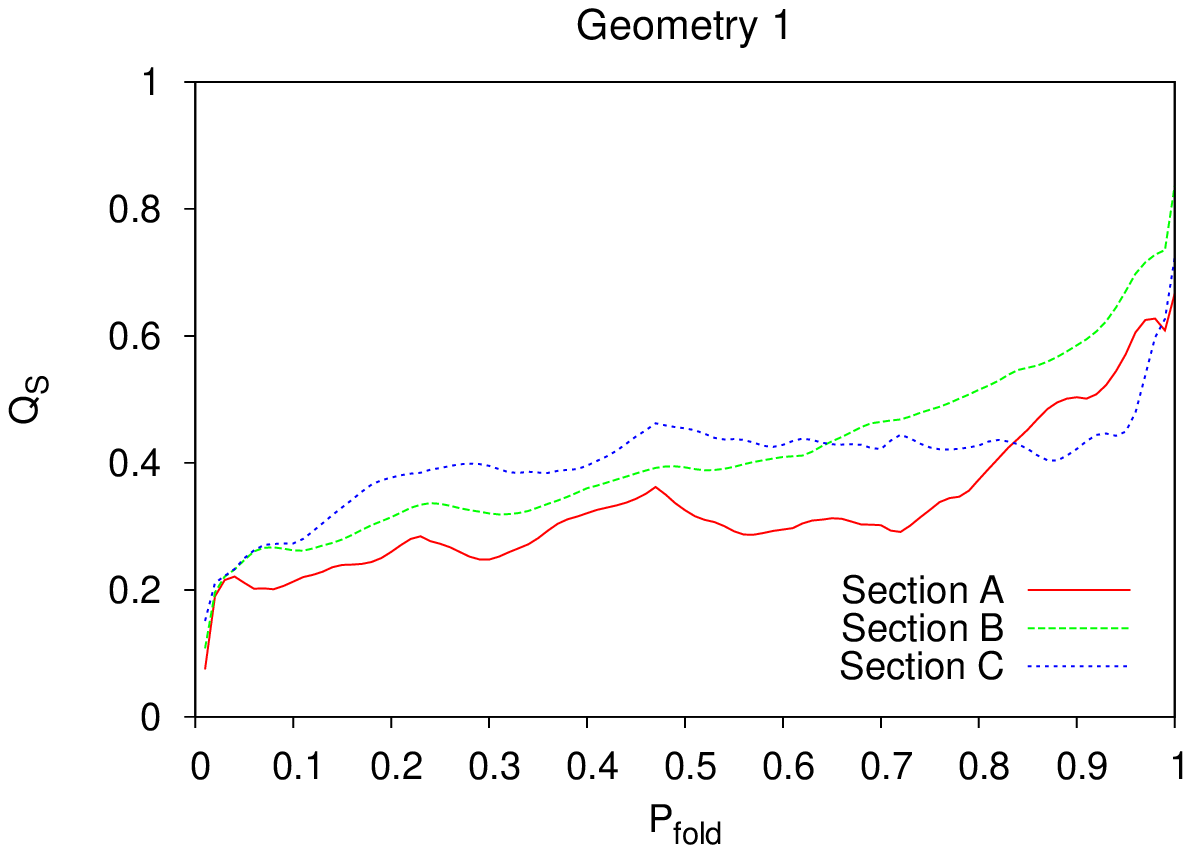, width=7.5cm} & \epsfig{file=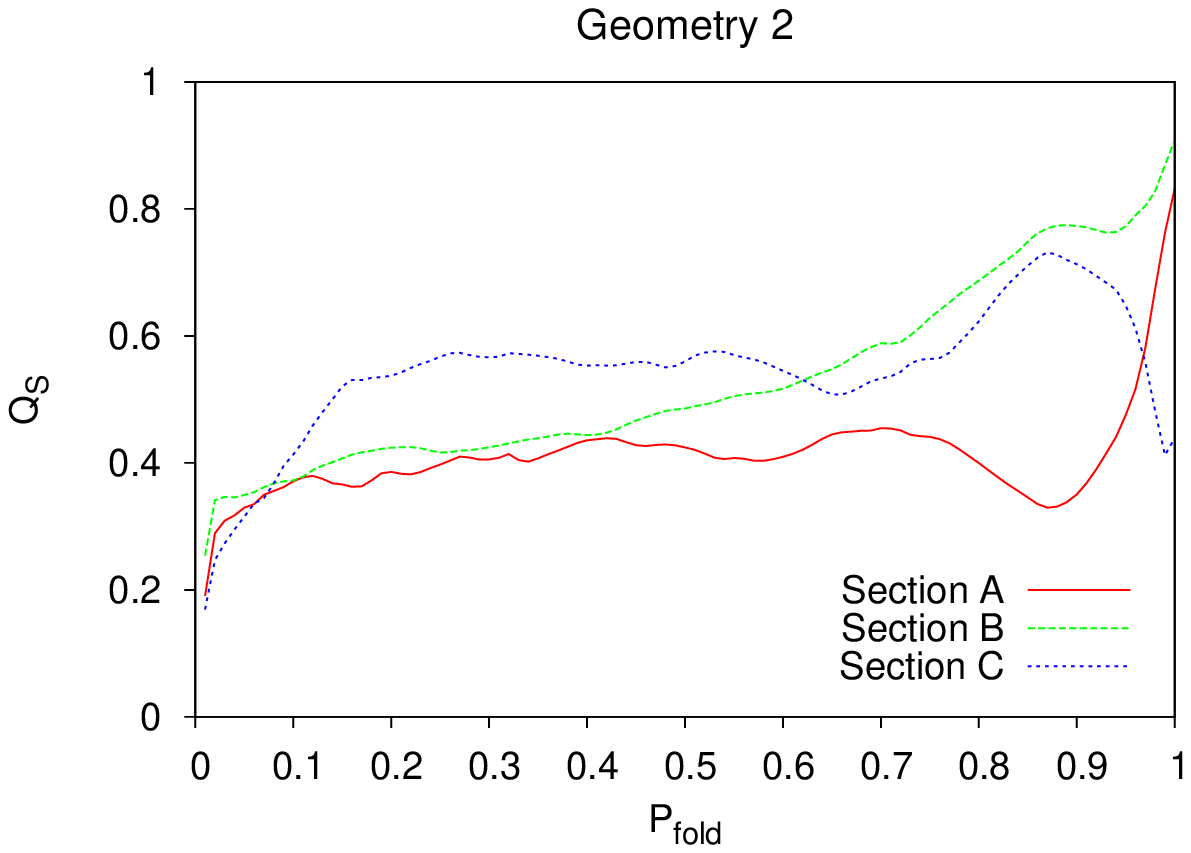, width=7.5cm}\\
\mathrm{(a)} & \mathrm{(b) }
\end{array}
$$
$$
\begin{array}{c}
\epsfig{file=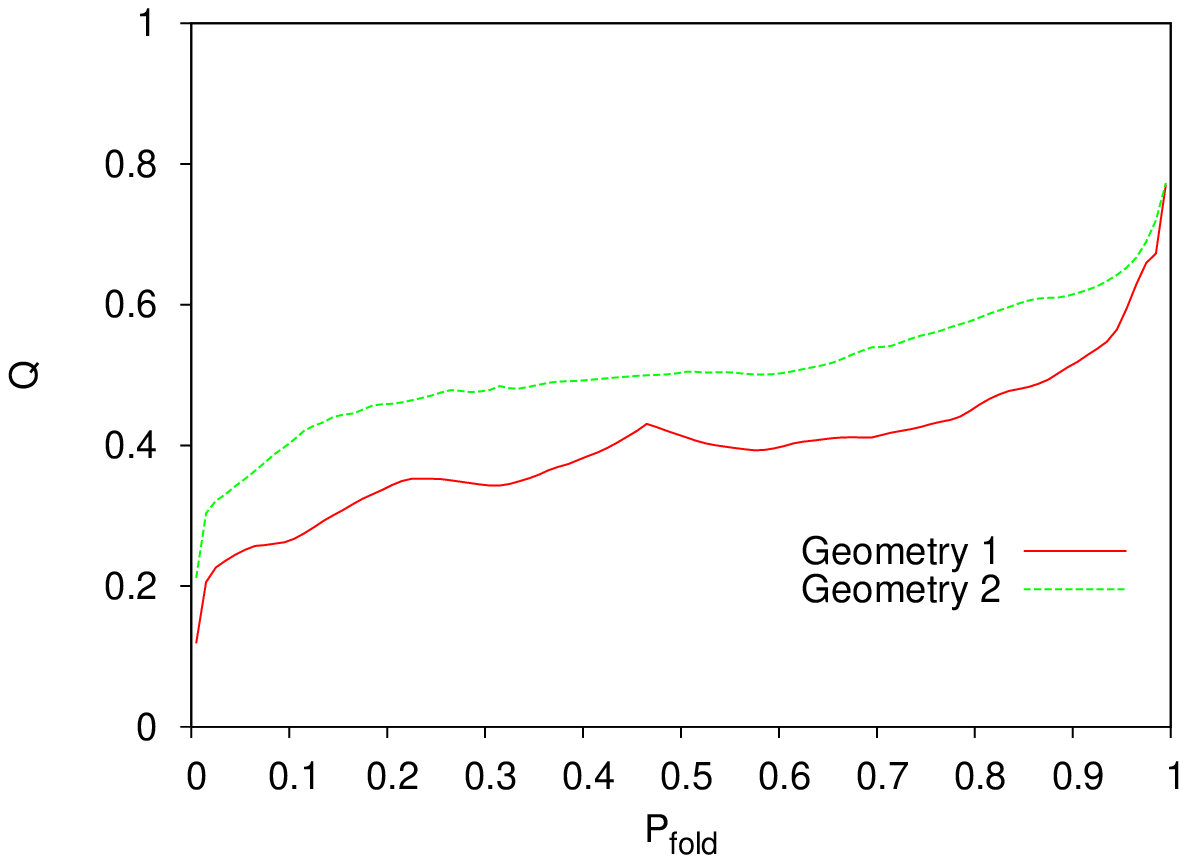, width=7.5cm}\\
\mathrm{(c) }
\end{array}
$$
\caption{Average fraction of native bonds in each protein section, $Q_s$, as a function of
$P_{fold}$ in Geometry 1 (a) and Geometry 2 (b). Also shown is the dependence of the 
protein's average fraction of native bonds on the reaction coordinate, $P_{fold}$,
for both geometries. Note that when folding is near completion at high $P_{fold}$ there
is a sharp increase in the fraction of native contacts for Geometry 1.    
 \label{junto}} 
\end{figure}

To gain further insight into the folding reactions of both model proteins 
we have determined how the average fraction of native bonds in each section, 
$<Q_S>$, changes with $P_{fold}$ (Figure \ref{junto}). 

The average fraction of native bonds in sections A and C decreases considerably 
when folding of Geometry 2 is near completion (Figure \ref{junto}(b)), which is    
suggestive of existing unfolding events at large $P_{fold}$. This presumably
happens due to a partial or complete folding of both section A and C prior to the 
complete folding of section B. Such unfolding events, which are required to 
ensure that folding follows the right pathway, do not occur to such an 
extent in Geometry 1 where A and B cannot fold at low $P_{fold}$ (Figure \ref{junto}(a)), perhaps due to topological constraints. A comparison of the data 
reported on Figures \ref{junto}(a) and \ref{junto}(b) with that shown in Figure \ref{junto}(c) 
indicates that the folding of the whole protein follows the folding of section B, in agreement
with the idea that section B drives the folding of the whole native structure.

\section{From macro- to micro-structural formation: Evidence for nucleation 
phenomena}

A post-critical folding nucleus (FN) is defined as a set of native bonds which, 
once formed, prompts rapid and highly probable folding~\cite{SHAKHNOVICH}.
We have recently developed a methodology, based on the concept of folding 
probability, aimed at identifying critical (i.e. nucleating) bonds in the 
folding of small lattice proteins~\cite{jphyscondensmat}. In a related effort, 
a simulational proxy of the phi-value analysis was used to identify nucleating
residues in the folding of the two model proteins investigated in the present work
~\cite{FAISCA2007}. We have found that the set of residues 6, 33 and 35 
in Geometry 1, and residues 19, 20 and 29 in Geometry 2 lead to the largest 
increase in folding time upon mutation. Interestingly, the vast majority 
(i.e. more than 60\%) of contacts formed by these residues are present in
section B of both proteins.\par 
The conclusion that section B encapsulates the nucleating residues (and therefore the 
set of native bonds forming a post-critical FN) can actually be drawn from an independent analysis of the results obtained so far. In addition to shedding light on the existence of protein sections, the contact correlation analysis introduced in section \ref{secant}, 
shows that the folding of section B is a pre-requisite to observe
inevitable (i.e. highly probable) folding of the whole protein.
Indeed, section B is always folded in the high-$P_{fold}$ conformations that 
are on-pathway to the native state (i.e. that fold fast). Therefore, if these model 
proteins fold via nucleation, section B must necessarily contain the critical residues 
forming the FN. In support of this argument we have found an increase in the correlation between $Q_{B}$ and $P_{fold}$ for both geometries when folding is near completion (i.e. $P_{fold} > 0.85$), which implies that the folding of section B determines the inevitable folding of the whole protein. 
For example, in Geometry 1, a conformation where section B is folded has folding probability $P_{fold} > 0.93$. 
Also illuminating is the fact that the presence, with high probability, of the bonds forming 
section B is independent of which other bonds are formed in the protein. Moreover, the most probable folding pathways are those in which section B folds early, while those in which it 
folds last have a vanishingly small probability of occurrence.

\section{Cooperativity at the level of macro-structural formation}

In protein folding the term cooperativity is generally used in 
connection with specific thermodynamic and kinetic features
exhibited by small, single domain proteins. Indeed, extraordinary 
experimental traits such as the linear chevron behavior 
(kinetic cooperativity), and the verification of the van t'Hoff 
criterion (thermodynamic cooperativity) have been typically ascribed 
to the existence of highly unusual energetics involving non-additive 
multi-body interactions~\cite{CHANREVIEW, FAISCAPLAXCO}.\par      
The results reported in this work are suggestive that Geometry 1 folds in a more
cooperative manner than Geometry 2. This difference in cooperative behavior is particularly evident from the study of section formation along the reaction coordinate (Figure \ref{sqq}). Here, it is shown that both section A and C in Geometry 2, can  have the vast majority of its bonds formed ($Q_{S}>0.9$) early on in folding (i.e. at low $P_{fold}$). In Geometry 1, on the other hand, the formation of bonds within one section does not happen in such an independent manner. Indeed, it is only when folding of the overall protein is near completion (i.e. for $P_{fold}>0.9$) that the fraction of bonds within each section comes close to unity. Also suggestive of the more cooperative behavior of Geometry 1 are the considerably smaller times elapsing between the formation of the first section and the folding of the whole protein (\ref{sfo}). Indeed, not only these time intervals are considerably smaller in Geometry 1 than in Geometry 2, as they are (on average) 33\% smaller than the cut-off time that was used to select the conformations that fold inevitably fast from other high $P_{fold}$ conformations (section \ref{sect}). For Geometry 2 such time intervals are similar to this cut-off parameter and much larger than the average folding time of conformations on pathway with $P_{fold}>0.9$. 
These times are in line with the finding that the first section to fold can do it relatively early 
during the process (i.e. $P_{fold} < < 0.9$), 
Finally, the higher cooperativity of Geometry 1 is also evident from the sharper increase in the fraction of native contacts, $Q$, that is observed near the very end of its folding 
process (Figure \ref{junto}(c)).

\section{Conclusions}

In the present work we investigated the existence of folding pathways
for two model proteins differing in native geometry at a coarse-grained 
level of structure formation. To this end we developed and applied  a methodology, based on 
native contact correlation analysis, which identifies 
protein sections with clusters of highly correlated native bonds. The latter
were shown to map onto well defined structural three dimensional domains within the native fold.\par  

Three protein sections and four folding pathways, 
corresponding to different ordering preferences of section formation, were 
identified for each protein. Interestingly, the analysis of folding pathways at
a macro-structural level of structure formation revealed a common underlying 
folding mechanism, based on nucleation phenomena, for both target 
geometries. Indeed, our results show that one of the protein
sections contains a set of critical bonds that form the folding nucleus.  
In the most complex geometry this section, and the folding nucleus, have a 
topology similar to that of the native fold~\cite{FERSHT, PACI}. 
  
Despite these similarities, it was identified a relevant difference between the folding
processes related to the different cooperative
behavior of the two proteins. The higher cooperativity observed for the 
most complex geometry is probably due to the larger number of non-local, 
long-range native bonds of the native fold as well as of the folding 
nucleus~\cite{GO}.
In other words the higher cooperativity of the folding process of the 
complex geometry is ascribed to the non-trivial order of the native fold, 
that is mimicked by that of of the folding nucleus. Despite the small size 
of the two model proteins this structural difference has a marked effect 
in the dynamics of the folding process and for the complex geometry it 
resembles the dynamics of first order transitions in the thermodynamic
limit. Quantitative measures of cooperativity, and in particular 
the size dependence of the nucleation barrier for the different geometries,
are outside the scope of this work~\cite{FRENKEL}.

We speculate that by introducing chemical specificity in our model proteins
the number, or at least the probability of occurrence, of the folding pathways identified
here, and that are solely driven by native geometry, will probably change. 
The use of a sequence-specific model (e.g. using the Miyazawa-Jernigan potential) 
is, however, out of the scope of the present study and will be investigated in future work.

\section{\bf {Acknowledgments}}R.D.M.T. and P.F.N.F. thanks Funda\c c\~ao para a Ci\^encia e Tecnologia (FCT) for financial support through grants SFRH/BPD/26093/2005 and  SFRH/BPD/21492/2005 respectively. This work was also supported by FCT through projects POCI/FIS/55592/2005 and POCTI/ISFL/2/618.

\end{document}